\documentclass[aps,pre,singlecolumn,superscriptaddress]{revtex4}

\usepackage{amsmath, amsfonts}
\usepackage{graphicx} 
\usepackage{physics}  
\usepackage{color}    
\usepackage{tikz}     
\usepackage{comment}
\usepackage{subcaption} 
\usepackage{float}    
\usepackage{mathtools} 
\usepackage[english]{babel} 
\usetikzlibrary{arrows.meta, calc, shapes.misc}

\newcommand{\beq}{\begin{equation}}
\newcommand{\eeq}{\end{equation}}
\newcommand{\eps}{\varepsilon}

\begin{document}

\title{Exponential Asymptotics for Dark Solitons of the Discrete NLS Model}

\author{C. J. Lustri}
\affiliation{School of Mathematics and Statistics, The University of Sydney, Camperdown 2050, Australia}

\author{P. G. Kevrekidis}
\affiliation{Department of Mathematics and Statistics, University of Massachusetts Amherst, Amherst, MA 01003, USA}

\affiliation{Department of Physics, University of Massachusetts Amherst, Amherst, MA 01003, USA}

\affiliation{Department of Mechanical Engineering, Seoul National University, 1 Gwanak-ro, Gwanak-gu, Seoul 08826, South Korea}

\author{D.E. Pelinovsky}
\affiliation{Department of Mathematics,
McMaster University, Hamilton, Ontario, Canada}

\begin{abstract}
In the present work we revisit the problem of
the dark solitary wave pinned in the discrete nonlinear
Schr{\"o}dinger equation. In a number of recent
studies, the methodology of exponential asymptotics
was attempted to be utilized in this problem, however
the results were not found to be fully in agreement
with associated multiprecision numerical 
computations. Here we resolve this conundrum 
by finding precise exponential asymptotics for the pinned dark 
solitary waves. Moreover, we reconcile
the relevant result with a general theory of
pinned dark solitary waves in the {\it continuum} nonlinear Schr{\"o}dinger equations 
in the presence of external potentials.
\end{abstract}

\maketitle

\section{Introduction}

The study of discrete nonlinear dispersive systems and of the effects
of the underlying lattice on the dynamics of solitary
waves has a long history of over half a century~\cite{PEYRARD198488}
that has by now been summarized in numerous reviews~\cite{chriseil,Flach2008},
as well as books~\cite{kev09,Pelinovsky_2011}. The relevant features
concern the existence of the famous Peierls-Nabarro 
barrier~\cite{campbell}, its dynamical implications 
on the mobility of kinks~\cite{PEYRARD198488} and 
discrete solitary waves~\cite{GOMEZGARDENES2004213,oxtoby},
the existence of discrete breathers~\cite{RSMacKay_1994,Flach2008}
(and their mobility~\cite{Flach1999}), as well as the existence
of discrete vortices~\cite{cretegny,JOHANSSON1998115,DV}, among
many others. 

A key feature of discreteness is that it breaks continuous symmetries,
such as most notably translational invariance, but also potentially
others such as the so-called conformal (scaling) invariance of the 
2d nonlinear Schr{\"o}dinger (NLS) model~\cite{KEVREKIDIS2012982}. 
This paves the way for approaching the continuum limit in different ways. For instance,
even within the same continuum model, such as, e.g., the NLS or 
sine-Gordon, and for a single discretization of such a model
respectively ---e.g., the discrete NLS (DNLS)~\cite{kev09} or
the discrete sG~\cite{PEYRARD198488}--- there exist multiple
stationary discrete solitary waves that correspond to the same
continuum solitary wave. I.e., there is an onsite (site-centered)
and an intersite (intersite-centered) version of a solitary wave,
one of which turns out to be spectrally stable and the other one 
turns out to be spectrally unstable, both of which approach the
same continuum solitary wave limit as the spacing distance between
lattice nodes tends to 0. However, additionally~\cite{kev09,KEVREKIDIS2012982},
even for the same continuum model, there may exist 
{\it multiple discretizations} (in the same way in which we envision
multiple discretizations in numerical analysis) which may preserve
(or ``destroy'') different symmetries. A characteristic example
at the NLS level is ---in addition to the standard DNLS discretization---
the famous integrable Ablowitz--Ladik discretization~\cite{AblowitzPrinariTrubatch}. 
The latter preserves infinitely many conservation laws and has,
accordingly, continuous families of solitary waves compared to the onsite and intersite states in the DNLS model.

In the present work we revisit one of the quintessential features that
discreteness {\it generically} introduces in nonlinear dynamical systems,
namely the implications of translation invariance breaking to the 
eigenvalues of the spectral stability problem for the discrete solitary waves as the continuum
limit is approached. This is a topic that has been explored
extensively in the DNLS model for at least a quarter of a century,
both at the level of bright solitary waves (of the focusing DNLS)~\cite{johaub}, as well as for dark solitary waves (of the defocusing
DNLS)~\cite{johkiv}. It is important to appreciate here that part
of the interest in the subject stems from the fact that this
is not a feature particular to this model but the breaking
of translational invariance will {\it very broadly} bear
such implications in discrete models, unless the models
are constructed in a way such that they preserve
translational invariance~\cite{Pelinovsky_2006}.
It was recognized already quite early (and was accordingly
quantified~\cite{Todd_Kapitula_2001}) that the breaking of
translational invariance should be beyond all algebraic orders
in the expansion of a discrete difference operator
in continuum derivatives. Indeed, such early works as~\cite{Todd_Kapitula_2001} were able to capture the exponentially
small behavior of associated eigenvalues to the continuum
limit (as well as the corresponding power law ``modulating'' this 
exponential dependence). However, for technical reasons related to the
methods used, such approaches were unable to identify the
relevant prefactor (or for that matter the correction to
the leading order exponential asymptotics).

Remarkably, recent efforts have brought to bear advanced 
{\it exponential asymptotics} techniques to bridge this gap
and provide definitive closure to the dependence of the relevant
eigenvalues both as regards the exponential and power-law dependences of the
prefactor on the lattice spacing, but also of the numerical prefactor~\cite{ADRIANO2025134848,LustriKevrekidisChapman2025}, as well as of
the leading order correction~\cite{LustriKevrekidisChapman2025}.
Indeed, relevant techniques were extended to other problems, such
as, e.g., ones involving next-nearest-neighbors~\cite{lustri2025borelpadeexponentialasymptoticsdiscrete}.
Importantly, also, concurrent developments in numerical computation
have enabled the identification of relevant eigenvalues with
unprecedented accuracy, enabling their monitoring over
{\it many} decades of data, which, in turn, has allowed to test
the accuracy of both the leading order predictions, but also
that of the associated next-order corrections~\cite{KusdiantaraAdrianoSusanto2025}.

Interestingly, the above developments can now be considered
to be {\it definite} (as described in the above paragraph)
{\it solely} for the case of bright solitons of the focusing 
NLS model that asymptote to a vanishing background. The realm of 
{dark solitons} of the defocusing DNLS model ---asymptoting
to two opposite non-vanishing constants--- has proven to
be far more elusive. Here, comparison of theoretical
attempts and numerical findings has been {\it unsuccessful}~\cite{ADRIANO2025134848} and even the most recent numerical
efforts~\cite{KusdiantaraAdrianoSusanto2025} have at best only
provided numerical fits to the eigenvalue data. The dark 
soliton case, as was already shown in the seminal work
of~\cite{johkiv} is far more technically involved as the continuous
spectrum of these solitary waves encompasses the
entire imaginary axis as the continuum limit is approached. Among the two cases of the onsite
and intersite discrete solitary waves, the one that is easier
to numerically monitor is the intersite one, involving always
a real eigenvalue pair, immediately upon departure from the well-known
anti-continuum limit~\cite{RSMacKay_1994,Pelinovsky_2008} of uncoupled
sites
to the continuum limit of vanishing spacing (and effectively
infinite coupling) between them. Even for that case, and despite
the most recent highest accuracy numerics~\cite{KusdiantaraAdrianoSusanto2025}, no theory has
captured the numerical prefactor (or the next-order-correction)
to this date, to the best of our knowledge. Far more subtle
is the case of the onsite mode where the translational eigenvalue
starts on the imaginary axis in the anti-continuum limit \cite{Pelinovsky_2008} and
upon a collision with the continuous spectrum that was 
definitively displayed in~\cite{johkiv}, becomes a {\it quartet} of eigenvalues
in the complex plane.
The relevant complex eigenvalue is exceptionally  difficult
to technically capture as has been illustrated not only in the seminal efforts 
of~\cite{johkiv}, but even within the most recent study of~\cite{KusdiantaraAdrianoSusanto2025}
(which has aimed to fit this eigenvalue only up to $h \approx 0.9$).

The aim of the present work is to {\it definitively} address
this issue through the use of delicate exponential asymptotics
that account effectively also for the role of
the continuous spectrum. Our efforts are significantly inspired
from the work of~\cite{PelinovskyKevrekidis2008}, which, in addition
to a Melnikov analysis for the persistence conditions of 
dark solitons in the perturbed continuum NLS equation, had developed
a stability analysis of the pinned dark solitary waves that also had
the same pathological characteristic of the splitting of small eigenvalues due to
an external potential breaking translational invariance and due to 
the continuous spectrum which admits no spectral gap near the origin. 
In that case too, the eigenvalues
when not directly real would move into the complex plane,
as we find here too. Indeed, we are able to identify both the
leading order prefactor of the real eigenvalues for the intersite
discrete soliton case, and the corresponding correction term.
Similarly, in the onsite case, we are able to trace the
complex eigenvalues and we come as close as we have found
it to be numerically possible to identifying the associated 
real and imaginary parts. In addition to the state-of-the-art
analytical findings, multiprecision numerical techniques have
been used to obtain many decades of numerical eigenvalues,
allowing the definitive testing of our findings.

Our presentation is structured as follows. Section II presents
the model setup and the existence analysis for the discrete dark 
solitary waves with the account of the exponential small terms. 
Section III lies at the center of our exposition and presents
our stability results. Section IV corroborates our findings
through detailed numerical computations of both intersite
and onsite eigenvalue configurations. Finally, section V
summarizes our findings and presents our conclusions.

\section{Mathematical Setup and Existence Theory}

Starting with the standard DNLS model~\cite{kev09}
\begin{eqnarray}
    i \dot{u}_n =-C (u_{n+1} + u_{n-1} -2 u_n) + |u_n|^2 u_n,
    \label{eqn0}
\end{eqnarray}
we consider, as is common within this model, the case of standing waves of the
form (setting the background amplitude/frequency to unity without loss of generality)
$u_n=e^{-i t} \phi_n$. Also, in order to connect with the continuum limit we
use $C=1/h^2$, aiming to characterize the limit of $h \rightarrow 0$,
rather than the anti-continuum limit of $C \rightarrow 0$.

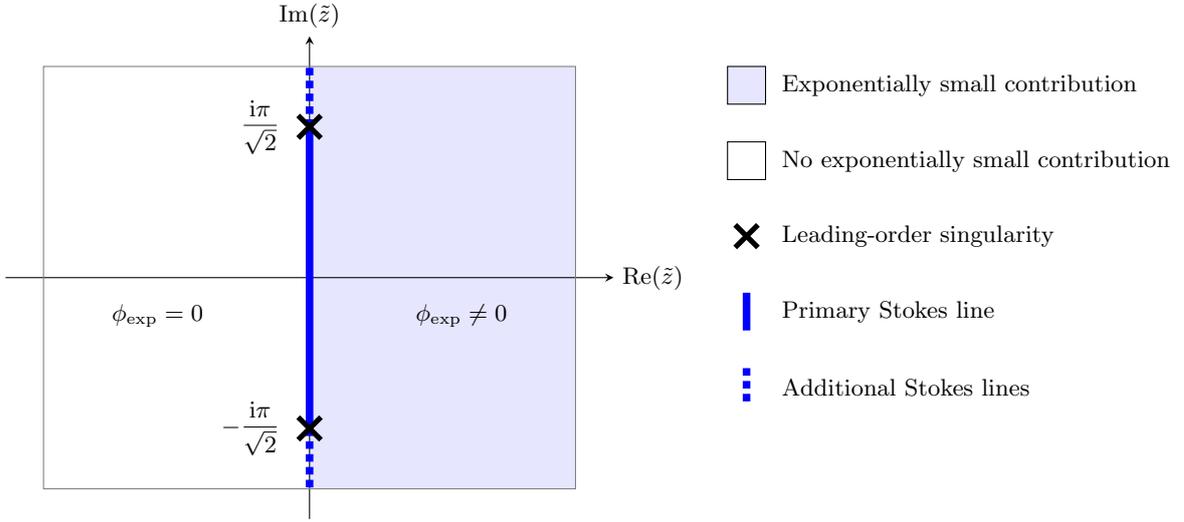
\begin{figure}[tb]
\centering
\begin{tikzpicture}
[xscale=1,>=stealth,yscale=1]

\fill[blue!10] (0,-2.8) -- (3.5,-2.8) -- (3.5,2.8) -- (0,2.8) -- cycle;
\draw[->] (-4,0) -- (4,0) node[right] {{$\mathrm{Re}(\tilde{z})$}};
\draw[->] (0,-3.2) -- (0,3.2) node[above] {{$\mathrm{Im}(\tilde{z})$}};
\draw[blue,line width=1mm] (0,-2) -- (0,2);
\draw[blue,line width=1mm,dotted] (0,-2) -- (0,-2.8);
\draw[blue,line width=1mm,dotted] (0,2) -- (0,2.8);
\draw[line width=0.7mm] (-0.15,1.85) -- (0.15,2.15);
\draw[line width=0.7mm] (-0.15,2.15) -- (0.15,1.85);
\draw[line width=0.7mm] (-0.15,-1.85) -- (0.15,-2.15);
\draw[line width=0.7mm] (-0.15,-2.15) -- (0.15,-1.85);
\node[left] at (-0.275,2) {$\dfrac{{\rm i}\pi}{\sqrt{2}}$} ;
\node[left] at (-0.275,-2) {$-\dfrac{{\rm i}\pi}{\sqrt{2}}$} ;

\node at (-2,-0.5) {{$\phi_{\mathrm{exp}} = 0$}};
\node at (2,-0.5) {{$\phi_{\mathrm{exp}} \neq 0$}};

\fill[blue!10] (5.5,2.3) -- (5.5,2.8) -- (6,2.8) -- (6,2.3) -- cycle;
\draw (5.5,2.3) -- (5.5,2.8) -- (6,2.8) -- (6,2.3) -- cycle;
\draw (5.5,1.3) -- (5.5,1.8) -- (6,1.8) -- (6,1.3) -- cycle;
\draw[line width=0.7mm] (5.6,0.7) -- (5.9,0.4);
\draw[line width=0.7mm] (5.6,0.4) -- (5.9,0.7);
\draw[blue,line width=1mm] (5.75,-0.2) -- (5.75,-0.7);
\draw[blue,line width=1mm,dotted] (5.75,-1.2) -- (5.75,-1.7);
\node[right] at (6.1,2.55) {Exponentially small contribution};
\node[right] at (6.1,1.55) {No exponentially small contribution};
\node[right] at (6.1,0.55) {Leading-order singularity};
\node[right] at (6.1,-0.45) {Primary Stokes line};
\node[right] at (6.1,-1.45) {Additional Stokes lines};

\draw[gray] (3.5,2.8) -- (3.5,-2.8) -- (-3.5,-2.8) -- (-3.5,2.8) -- cycle;
\end{tikzpicture}
\caption{Stokes structure for the discrete 
solitary wave solution. The relevant Stokes line is shown as a solid blue line. There are additional singularities further along the imaginary axis in both directions, but the corresponding Stokes lines switch on asymptotic contributions that are exponentially small compared to those considered here.}\label{fig:stokes}
\end{figure}

The governing steady state model accordingly reads: 
\begin{equation}
\label{discrete-nls}
    \frac{1}{ h ^2}(\phi_{n+1} - 2 \phi_n + \phi_{n-1}) - (|\phi_n|^2-1)\phi_n = 0.
\end{equation}
Let $z =  h  n$ and $\tilde z = z-z_{0}$ with $z_0 \in \mathbb{R}$ to be defined. Expanding Eq. (\ref{discrete-nls}) in powers of $ h ^2$ gives 
\begin{equation}
\label{lead-order}
2\sum_{m=1}^{\infty}\frac{ h ^{2m-2}}{(2m)!} \frac{\mathrm{d}^{2j}\phi(\tilde{z})}{\mathrm{d}\tilde{z}^{2j}} + (\phi(\tilde{z})^2 - 1)\phi(\tilde{z}) = 0,
\end{equation}
where we have made the assumption that $\phi(\tilde{z})$ is real-valued for real $\tilde{z}$. 
We are looking for the power series solution of Eq. (\ref{lead-order}) given by
\begin{equation}
    \phi(\tilde{z}) \sim \sum_{j=0}^{\infty} h ^{2j}\phi_j(\tilde{z}) \quad \mathrm{as} \quad  h  \to 0,
\end{equation}
where we have abused notation slightly by using subscripts to indicate the series index, as opposed to the value of the lattice index $n$, for the sake of later notational convenience. We select the leading-order solution that tends to the dark solitary wave in the continuum limit,
\begin{equation}\label{e:phi0}
\phi_0(\tilde z)=\tanh\!\left(\frac{\tilde z}{\sqrt{2}}\right).
\end{equation}
This leading-order solution is singular at $\tilde{z}_s = {\mathrm{i}\pi}/{\sqrt{2}} + k \sqrt{2}\pi \mathrm{i}$ for $k \in \mathbb{Z}$. The important singularities are $k = 0$ and $-1$, which are located at complex conjugate positions. The recurrence relation for the series terms is
Expanding this gives
\begin{equation}
       2\sum_{m=1}^{j}\frac{1}{(2m)!} \frac{\mathrm{d}^{2j}\phi_{j-m+1}}{\mathrm{d}\tilde{z}^{2j}} + (3\phi_0^2 - 1)\phi_j = 0.
\end{equation}
We can use the recurrence relation to obtain the particular solution for $\phi_1(\tilde{z})$, given by
\begin{equation}\label{e:phi1}
    \phi_1(\tilde{z}) = -\frac{1}{48}\sech^2\left( \frac{\tilde z}{\sqrt{2}} \right)\left(9 + 2\sqrt{2}\tilde{z} - 8 \tanh\left( \frac{\tilde z}{\sqrt{2}} \right) \right).
\end{equation}
This will allow us to find the first correction in $ h $ to the eigenvalue calculation.
The late-order behaviour has the form
\begin{equation}\label{e:ansatz}
    \phi_j(\tilde z) \sim \frac{\Phi(\tilde{z})\Gamma(2j + \gamma)}{\chi(\tilde{z})^{2j + \gamma}}\quad \mathrm{as} \quad n \to \infty,
\end{equation}
where $\chi(\tilde{z}) = 0$ at $\tilde{z} = \tilde{z}_s$. Using the same techniques as \cite{LustriKevrekidisChapman2025}, we find that the singulant is given by
\begin{equation}
    \chi(\tilde{z}) = 2\pi\mathrm{i}\ell (\tilde{z} - \tilde{z}_s),\qquad \ell \in \mathbb{Z},
\end{equation}
where we are concentrating on $\ell = \pm 1$. This means that the Stokes lines emerge from the singularity at $\tilde{z} = \tilde{z}_s$ and follow the imaginary axis, as in \cite{LustriKevrekidisChapman2025}. This is shown in Figure \ref{fig:stokes}. The prefactor equation becomes
\begin{equation}
    \Phi'' - (3 \phi_0^2 - 1)\Phi = 0,
\end{equation}
so $\Phi(\tilde{z}) = K_1 \Phi_1(\tilde{z}) + K_2 \Phi_2(\tilde{z}) $, where $K_1$ and $K_2$ are arbitrary constants, while
\begin{align}\label{e:Phi1}
\Phi_{1}(\tilde z)&=\frac14\,\operatorname{sech}^{2}\left( \frac{\tilde z}{\sqrt{2}} \right),\\
\Phi_{2}(\tilde z)&=\frac12\Bigl(6\sqrt{2}\,\tilde z+8\sinh(\sqrt{2}\tilde z)+\sinh(2\sqrt{2}\tilde z)\Bigr)\,
\operatorname{sech}^{2}\left( \frac{\tilde z}{\sqrt{2}} \right).
\label{e:Phi2}
\end{align}
This choice of homogeneous solution is so that $\Phi_1 \sim \mathrm{e}^{-\sqrt{2}\tilde z}$ as $\tilde z \to \pm\infty$ and $\Phi_{2}\sim \pm \mathrm{e}^{\sqrt{2}\tilde z}$ as $\tilde z\to \pm\infty$. The parameters $\gamma$ and $K_1$ must be determined by considering the late-order terms \eqref{e:ansatz} in the vicinity of $\tilde{z} = \tilde{z}_s$. By using an essentially identical argument to \cite{LustriKevrekidisChapman2025}, it follows that $\gamma = 4$ and that $K_1 \approx -5\mathrm{i}\Lambda/(4\sqrt{2})$ where $\Lambda \approx 358.464$.

Using a largely unchanged exponential asymptotic analysis in comparison to \cite{LustriKevrekidisChapman2025}, which was based on the method introduced in \cite{chapman2005exponential,king_chapman_2001,Daalhuis}, it can therefore be shown that the exponentially-small quantity which appears as the Stokes line, shown in Figure \ref{fig:stokes}, is crossed from left to right is given by
\begin{equation}\label{e:phiexp}
\phi_{\exp}(\tilde{z}) \sim -\frac{2\pi\mathrm{i} \Phi(\tilde z)}{ h ^{\gamma}}\,e^{-\frac{\chi(\tilde{z})}{ h }}\,
+ \textrm{c. c.} \quad \mathrm{as} \quad  h  \to 0,
\end{equation}
where c. c. refers to the complex conjugate of this expression. In the limit as $\tilde{z} \to \infty$, this is
\begin{equation}\label{e:phiexpinf}
\phi_{\exp}(\tilde{z}) \sim -\frac{2\pi}{ h ^{4}}\left(\frac{5\Lambda}{4\sqrt{2}}\mathrm{e}^{\sqrt{2}\tilde{z}}\right)\,e^{-\frac{\sqrt{2}\pi^2}{h} }\,
\sin\!\left(\frac{2\pi \tilde z}{ h }\right) \quad \mathrm{as} \quad  h  \to 0,\, \tilde{z} \to \infty.
\end{equation}
The exponential grows in the limit that $\tilde{z} \to \infty$, which is inconsistent with the requirement that the solution remain bounded on the original discrete sites ($\epsilon z/h \in \mathbb{Z}$) in this limit.

As in the DLNS bright breather case, we can take advantage of the fact that we are free to select $z_0$ in order to identify solutions in which $\phi_{\mathrm{exp}} = 0$ on discrete sites. By selecting $z_0$ such that $z_0/h = N$ where $N \in \mathbb{Z}$, we find that
\begin{equation}
    \sin\left(\frac{2\pi \tilde{z}}{h}\right) = \sin\left(\frac{2 \pi z}{h} - \frac{2 \pi z_0}{h}\right)= \sin\left(\frac{2 \pi z}{h} - 2\pi N\right)  = \sin\left(\frac{2 \pi z}{h}\right),
\end{equation}
which equals 0 if $z/h \in \mathbb{Z}$. This corresponds to site-centered solutions. If we instead select $z_0$ such that $z_0/h = N+1/2$ where $N \in \mathbb{Z}$, we obtain
\begin{equation}
    \sin\left(\frac{2\pi \tilde{z}}{h}\right) = \sin\left(\frac{2 \pi z}{h} - \frac{2 \pi z_0}{h}\right) = \sin\left(\frac{2 \pi z}{h} - (2N  + 1)\pi\right) = \sin\left(\frac{2\pi z}{h} \right),
\end{equation}
which also equals 0 if $z/h \in \mathbb{Z}$. This corresponds to selecting intersite-centered solutions. For both of these configurations, $\phi_{\mathrm{exp}}(\tilde{z}) \to 0$ as $\tilde{z} \to \infty$ on the discrete sites, even though the continuous function $\phi_{\mathrm{exp}}(\tilde{z})$ demonstrates exponentially growing oscillations for other values of $\tilde{z}$ in this limit. 

\section{Stability Analysis}

Obtaining the leading-order approximation to the eigenvalues as in \cite{LustriKevrekidisChapman2025} indicates that the intersite-centered solutions should be unstable with a pair of real eigenvalues near the origin while the site-centered solutions are stable with a pair of purely imaginary eigenvalues near the origin. However, we know from \cite{PelinovskyKevrekidis2008} that, in the latter cae, the pair of eigenvalues splits off the purely imaginary axis 
as a complex quadruplet. To explain this splitting, we must find the first correction in the expansions for the small eigenvalues in small $h$. This analysis is proximal to that of \cite{PelinovskyKevrekidis2008} in some respects; however, there are key differences that are presented.

We introduce the usual perturbation
$\phi=\phi_{s}+f e^{\lambda t}+g^{*}e^{\lambda^{*}t}$, where $\phi_{s}$ is either an onsite or intersite-centered solution. Setting $v=f+g$ and $w=f-g$ gives
\begin{equation}\label{eq:L-L+}
L_{+}v=-\mathrm{i} \lambda w,\qquad L_{-}w=\mathrm{i}\lambda v,
\end{equation}
where
\begin{equation}
L_{+}=2\sum_{m=1}^{\infty}\frac{ h ^{2m-2}}{(2m)!}\frac{\mathrm{d}^{2m}}{\mathrm{d}\tilde z^{2m}}-(3\phi_{s}^{2}-1),\qquad
L_{-}=2\sum_{m=1}^{\infty}\frac{ h ^{2m-2}}{(2m)!}\frac{\mathrm{d}^{2m}}{\mathrm{d}\tilde z^{2m}}-(\phi_{s}^{2}-1).
\end{equation}
We expand the small eigenvalues bifurcating from the zero eigenvalue for small $h$ as 
\begin{equation}
\label{lambda-exp}
    \lambda \sim \eps^{1/2} \lambda_1 + \eps \lambda_2 + \ldots \quad \mathrm{as} \quad \eps \to 0,
\end{equation}
where $\eps$ is exponentially small in $h$, see the precise definition (\ref{def-eps}) below. 
Each $\lambda_j$ in the expansion (\ref{lambda-exp}) is also expanded in powers of $h$, so 
this expansion describes an asymptotic trans-series in the small parameter $h$ \cite{aniceto2019primer}. We will therefore write
\begin{equation}
    \lambda_j \sim \lambda_j^{(0)} + h \lambda_j^{(1)} + \ldots \quad \mathrm{as} \quad h \to 0,
\end{equation}
for $j = 1, 2, \ldots$. We will later find that the first correction to $\lambda_1^{(0)}$, or $\lambda_1^{(1)}$, is necessary for the asymptotic analysis to match with the numerical calculations. It is possible that $\lambda_j^{(k)}$ can depend on $h$, but we do require each term in the asymptotic series to be smaller than the preceding terms in the small-$h$ limit.

%This scaling for $\eps$ is chosen to be consistent with \cite{PelinovskyKevrekidis2008}. We will later select $\eps$ to capture the exponentially small eigenvalue scaling in the small-$ h $ limit. This choice, not required in the analysis from \cite{LustriKevrekidisChapman2025}, allows us to study exponential corrections to the eigenvalue itself. 

Similarly to (\ref{lambda-exp}), we expand $v$ and $w$ to obtain
\begin{equation}
\label{v-w-exp}
    v \sim v_0 + \eps^{1/2} v_1 + \eps v_2 + \eps^{3/2} v_3 +\ldots,\qquad w \sim w_0 + \eps^{1/2} w_1 + \eps w_2 + \eps^{3/2} w_3 +\ldots,
\end{equation}
in the limit that $ h $, and hence $\eps$, tends to $0$. Due to the translation invariance in the limit, we begin with
\begin{equation}
v_{0}=\frac{\mathrm{d}\phi_{s}}{\mathrm{d}\tilde z},\qquad w_{0}=0,
\end{equation}
which has the asymptotic behaviour
\begin{equation}\label{e:v0inf}
v_{0}\sim -\frac{5\Lambda\pi^{2}}{\sqrt{2} h ^{5}}\,e^{\sqrt{2}\tilde z}\,e^{-\frac{\sqrt{2}\pi^{2}}{h} }
\cos\!\left(\frac{2\pi \tilde z}{ h }\right)
\qquad \text{as }\tilde z\to\infty,\  h \to 0.
\end{equation}
This is the growth that needs to be balanced by the $\mathcal{O}(\eps^{1/2})$ term to obtain the leading-order eigenvalue behaviour. Note that 
the leading-order behaviour in \cite{PelinovskyKevrekidis2008} was balanced by an $\mathcal{O}(\eps)$ correction to $L_{\pm}$. However, 
in the setting of the discrete problem, we cannot apply a correction to $L_{\pm}$ in the same way, as the exponential tail is zero on all sites and the correction to $L_{\pm}$ is therefore zero. Instead, we will enforce the cancellation between $v_0$ and $v_2$ directly.

\subsection{Calculating $\lambda_1^{(0)}$}

We balance terms from \eqref{eq:L-L+}, \eqref{lambda-exp}, and \eqref{v-w-exp} at $O(\eps^{1/2})$ to obtain
\begin{equation}
L_{+}v_{1}=0,\qquad L_{-}w_{1}=\mathrm{i} \lambda_1 \frac{\mathrm{d}\phi_{s}}{\mathrm{d}\tilde z}.
\end{equation}
The first equation gives $v_1 = a_{11} \Phi_1 + a_{12} \Phi_2$, where $\Phi_1$ and $\Phi_2$ are given by \eqref{e:Phi1}--\eqref{e:Phi2}.
To prevent exponential growth as $\tilde{z} \to \infty$, we require $a_{12} = 0$. As $\Phi_1$ decays exponentially in the limit $\tilde{z} \to \pm\infty$, we set $a_{11} = 0$ without loss of generality. Balancing the second equation at leading order in $ h $, we find
\begin{equation}
\label{w-1-eq}
\frac{\mathrm{d}^{2}w_{1}}{\mathrm{d}\tilde z^{2}}-\Bigl(\phi_0^2-1\Bigr)w_{1}
= \frac{\mathrm{i}\lambda^{(0)}_1}{\sqrt{2}}\,\operatorname{sech}^{2}\left(\frac{\tilde z}{\sqrt{2}}\right).
\end{equation}
The general solution of this inhomogeneous equation is 
\begin{equation}
w_{1}=\frac{\mathrm{i}\lambda^{(0)}_1}{\sqrt{2}}\left(1 +b_{11}W_{1}+b_{12}W_{2}\right),
\end{equation}
where $W_{1}$ and $W_{2}$ are homogeneous solutions given by 
\begin{equation}
\label{W1W2}
W_{1}(\tilde z)=\phi_0(\tilde z),\qquad W_{2}(\tilde z)=\sqrt{2}-\tilde z\,\phi_0(\tilde z).
\end{equation}
In order for this to tend to a constant as $\tilde z\to \infty$, which is required to match with the general form given in 
Theorem 4.10 of \cite{PelinovskyKevrekidis2008}, we need $b_{12}=0$. Any other choice of $b_{12}$ will give linear growth 
as $\tilde{z} \to \infty$, which is not permitted in the dual limit $\eps \to 0$ and $\tilde{z} \to \infty$. Since we have 
no information about $b_{11}$, we set 
\begin{equation}
\label{w1explicitly}
w_{1}=\frac{\mathrm{i}\lambda^{(0)}_1}{\sqrt{2}}\left(1+b_{11}\phi_0\right).
\end{equation}
Balancing terms from  \eqref{eq:L-L+}, \eqref{lambda-exp}, and \eqref{v-w-exp} at $O(\eps)$ gives
\begin{equation}\label{e:v2w2}
L_{+}v_{2}=-\mathrm{i} \lambda_1 w_{1},\qquad L_{-}w_{2}=\mathrm{i}\lambda_2 v_0.
\end{equation}
Balancing the first equation from \eqref{e:v2w2} at leading order in $ h $ shows that $v_{2}$ satisfies
\begin{equation}
\frac{d^{2}v_{2}}{d\tilde z^{2}}-\Bigl(3\phi_0^2-1\Bigr)v_{2}
=\frac{\left(\lambda^{(0)}_1\right)^2}{\sqrt{2}} \left(  1+b_{11} \tanh\left(\frac{\tilde z}{\sqrt{2}}\right) \right).
\end{equation}
Solving this gives 
\begin{align}\nonumber
    v_2 = \frac{\lambda^{(0)}_1}{32\sqrt{2}}\sech^2\left(\frac{\tilde z}{\sqrt{2}}\right) \Bigg(8\lambda^{(0)}_1\cosh(\sqrt{2}\tilde{z}) + 2\lambda^{(0)}_1\cosh(2\sqrt{2}\tilde{z}) + b_{11}\sqrt{2}  \sinh(\sqrt{2}\tilde{z}) - 14\lambda^{(0)}_1 - 4 b_{11} \tilde{z}\Bigg)&\\
    + a_{21} \Phi_1 + a_{22} \Phi_2.
\end{align}
As before, we set $a_{21} = 0$ without loss of generality. The particular solution grows exponentially in the limits that $\tilde{z} \to \pm\infty$, which is inconsistent with the growth predicted by Theorem 4.10 of \cite{PelinovskyKevrekidis2008}. We may choose $a_{22}$ to eliminate exponential growth as $\tilde{z} \to -\infty$ by selecting 
\begin{equation}\label{e:a22old}
%    a_{22} = \frac{\lambda_1^2}{8\sqrt{2}}.
a_{22} = \frac{\sqrt{2}\left(\lambda^{(0)}_1\right)^2 - b_{11}\lambda^{(0)}_1}{16}.
\end{equation}
As $\tilde{z} \to \infty$, the asymptotic behaviour of $v_2$ is a growing exponential,
\begin{equation}\label{e:v2inf}
v_{2}\sim \frac{\left(\lambda^{(0)}_1\right)^2}{4\sqrt{2}}\,e^{\sqrt{2}\tilde z}\quad \text{as } \quad  h \to 0,\ \tilde z\to \infty.
\end{equation}
In order to eliminate the growing exponential as $\tilde{z} \to \infty$ from $v_0$, we require
\begin{equation}
    v_0 \sim  -\eps v_2  \qquad \mathrm{as} \qquad \tilde{z} \to \infty.
\end{equation}
Using the asymptotic behaviour of $v_0$ from \eqref{e:v0inf} and the behaviour of $v_2$ from \eqref{e:v2inf}, this gives 
\begin{equation}
\label{lambda-1}
    \eps \left(\lambda_1^{(0)}\right)^2 \sim -\frac{20 \pi^2 \Lambda}{ h ^5} \mathrm{e}^{-\frac{\sqrt{2}\pi^2}{h} } \cos(2\pi (n-n_0)),
\end{equation}
where we have written the result using the discrete variable $n$ to emphasise the sign difference between onsite-centered (integer $n_0$) and intersite-centered (half-integer $n_0$) solutions. We absorb the exponential scaling into $\eps$ by setting
\begin{equation}
\label{def-eps}
    \eps = \mathrm{e}^{-\frac{\sqrt{2}\pi^2}{h} }.
\end{equation}
Recalling that $\lambda \sim \eps^{1/2} \lambda^{(0)}_1$ as $h \to 0$, the intersite-centered solutions have
\begin{equation}
\lambda \sim \eps^{1/2} \lambda^{(0)}_1 =  \pm\frac{\pi\sqrt{20\Lambda}}{ h ^{5/2}} 
\mathrm{e}^{-\frac{\sqrt{2}\pi^2}{2h}} \approx \pm \frac{266.004}{ h ^{5/2}} \mathrm{e}^{-\frac{\sqrt{2}\pi^2}{2h}}\quad \mathrm{as} \quad  h  \to 0.
\end{equation}
Since $\lambda^{(0)}_1$ is real, the intersite-centered solutions are unstable. The onsite-centered solutions have
\begin{equation}
\lambda \sim \eps^{1/2} \lambda^{(0)}_1 = \pm \frac{\mathrm{i} \pi\sqrt{20\Lambda}}{ h ^{5/2}}\mathrm{e}^{-\frac{\sqrt{2}\pi^2}{2h}} \approx \pm \frac{266.004\mathrm{i}}{ h ^{5/2}} \mathrm{e}^{-\frac{\sqrt{2}\pi^2}{2h}}\quad \mathrm{as} \quad  h  \to 0.
\end{equation}

\subsection{Calculating $\lambda^{(1)}_1$}
As the series terms in the expansion for $\lambda$ are themselves also series in $ h $, we can calculate the correction to these terms in the small-$ h $ limit. This correction is typically significant on the scale at which the numerical calculations will take place. This calculation follows the same steps as in \cite{LustriKevrekidisChapman2025}, and uses on the expressions for $\phi_0$ and $\phi_1$ from  \eqref{e:phi0} and \eqref{e:phi1}. In the limit $\tilde{z} \to \tilde{z}_s$, these have the form
\begin{align}
\phi_0(\tilde{z})\sim  \frac{\sqrt{2}}{\tilde{z}-\tilde{z}_s} + \frac{1}{3\sqrt{2}}(\tilde{z} - \tilde{z}_s) + \ldots,\qquad \phi_1(\tilde{z})\sim    -\frac{\sqrt{2}}{3(\tilde{z}-\tilde{z}_s)^3} + \frac{9 + 2\mathrm{i}\pi}{24(\tilde{z} - \tilde{z}_s)^2} + \ldots.
\end{align}
Combining these expressions yields
\begin{equation}
    \phi \sim \frac{\sqrt{2}}{\tilde{z}-\tilde{z}_s}+ \frac{1}{3\sqrt{2}}(\tilde{z} - \tilde{z}_s) +  h ^2\left( -\frac{\sqrt{2}}{3(\tilde{z}-\tilde{z}_s)^3} + \frac{9 + 2\pi\mathrm{i}}{24(\tilde{z} - \tilde{z}_s)^2}\right)\quad \mathrm{as} \quad  h \to 0,\, \tilde{z} \to \tilde{z}_s.
\end{equation}
Writing this in terms of the inner variable $ h  \eta = \tilde{z} - \tilde{z}_s$ and letting $\hat{\phi}(\eta) =  h  \phi(\tilde{z})$ gives a series containing an $\mathcal{O}( h )$ correction term,
\begin{equation}
\hat{\phi}(\eta) \sim \left(\frac{\sqrt{2}}{\eta} - \frac{\sqrt{2}}{3\eta^3}\right) + \frac{(9 + 2\pi\mathrm{i}) h }{24\eta^2} + \mathcal{O}( h ^2)\quad \mathrm{as} \quad  h  \to 0.
\end{equation}
There is a distinction between this analysis and the work of \cite{king_chapman_2001,LustriKevrekidisChapman2025}, in that we will not select an alternative inner variable to eliminate the entire $\mathcal{O}( h )$ term, but rather only the imaginary part of the numerator. The remaining contributions are eventually canceled by the complex conjugate term in the exponential asymptotic analysis, and can be ignored in this step. We therefore define the alternative inner variable 
$$ 
h  \hat{\eta} = \tilde{z} - \tilde{z}_s - \mathrm{i} \frac{\pi h}{12\sqrt{2}},
$$
which eliminates the imaginary part of the $\mathcal{O}( h )$ term. This leaves
\begin{equation}
\hat{\phi}(\eta) \sim \left(\frac{\sqrt{2}}{\eta} - \frac{\sqrt{2}}{3\eta^3}\right) + \frac{9 h }{24\eta^2} + \mathcal{O}( h ^2)\quad \mathrm{as} \quad  h  \to 0.
\end{equation}
This captures the first correction to the imaginary part of the true pole location, which is important for the eigenvalue calculation. Using an identical argument to \cite{LustriKevrekidisChapman2025}, the adjusted exponentially small contribution based on \eqref{e:phiexp} becomes
\begin{equation}
    \phi_{\mathrm{exp}}(\tilde{z}) \sim  -\frac{2\pi\mathrm{i} \Phi(\tilde z)}{ h ^{\gamma}}\,
    \mathrm{e}^{-\frac{\sqrt{2}\pi^2}{h} }\mathrm{e}^{-  \frac{\pi^2 h}{24}}\,
+ \textrm{c. c.} \quad \mathrm{as} \quad  h  \to 0.
\end{equation}
In the dual limit $\tilde{z} \to \infty$ and $ h  \to 0$, this term has the behaviour
\begin{equation}
\phi_{\exp}(\tilde{z}) \sim -\frac{2\pi}{ h ^{4}}\left(\frac{5\Lambda}{4\sqrt{2}}
\mathrm{e}^{\sqrt{2}\tilde{z}}\right)\,e^{-\frac{\sqrt{2}\pi^2}{h} }\,
\left(1 - \frac{\pi^2 h }{12\sqrt{2}}\right)\sin\!\left(\frac{2\pi \tilde z}{ h }\right) \quad \mathrm{as} \quad  h  \to 0,\, \tilde{z} \to \infty,
\end{equation}
and hence we find that 
\begin{equation}
    \lambda_1^{(1)} = - \frac{\pi^2\lambda_1^{(0)}}{12\sqrt{2}}.
\end{equation}
Thus, the corrected eigenvalue at the order of $\eps^{1/2}$ is given by 
\begin{equation}
\label{lambda-intersite}
    \lambda \sim \eps^{1/2} \lambda^{(0)}_1 \left(1 - \frac{\pi^2 h }{12\sqrt{2}}\right) \approx \pm (266.004 - 154.700h)h^{-5/2}
    \mathrm{e}^{-\frac{\sqrt{2}\pi^2}{2h}} \quad \mathrm{as} \quad  h  \to 0
\end{equation}
for intersite-centered solutions, and 
\begin{equation}
\label{lambda-onsite}
    \lambda \sim \eps^{1/2} \lambda^{(0)}_1 \left(1 - \frac{\pi^2 h }{12\sqrt{2}}\right) \approx \pm \mathrm{i}(266.004 - 154.700h)h^{-5/2}
    \mathrm{e}^{-\frac{\sqrt{2}\pi^2}{2h}} \quad \mathrm{as} \quad  h  \to 0
\end{equation}
for onsite-centered solutions.

As in \cite{LustriKevrekidisChapman2025}, this predicts that the onsite-centered eigenvalues are imaginary at leading order in $\epsilon$. 
However, in order to show that these solutions are also unstable, we need to find the correction term $\lambda_2$, which is turned out to be real.
This will connect with the expectation based on numerical observations~\cite{johkiv,KusdiantaraAdrianoSusanto2025}
about the instability of these onsite solutions due to a complex eigenvalue quartet.

\subsection{Calculating $\lambda^{(0)}_2$}
Balancing the second equation from \eqref{e:v2w2} at leading order in $h$ gives the equation for $w_2$, 
\begin{equation}
\label{w2explicit}
    w_2 = \mathrm{i}\left(\frac{\lambda^{(0)}_2}{\sqrt{2}} + b_{21}W_1 + b_{22} W_2\right),
\end{equation}
where $W_1$ and $W_2$ are given by \eqref{W1W2}. The coefficient $b_{21}$ does not affect the present calculation. From the analysis in Theorem 4.10 of \cite{PelinovskyKevrekidis2008}, we know that the perturbation $w_2$ can grow with the rate of $\mathcal{O}(\tilde{z}\eps)$ as $\tilde{z} \to \infty$ and $\eps \to 0$. To specify $b_{22}$, we must match the expansion for $w$ against the corresponding large-$\tilde{z}$ result from \cite{PelinovskyKevrekidis2008}.

\subsubsection{Determining the constant $b_{22}$}

From (4.12) in \cite{PelinovskyKevrekidis2008}, we know the limiting behaviour of $w$ as $\tilde{z} \to \pm\infty$. Expressed in the present variables (noting that $c$ from \cite{PelinovskyKevrekidis2008} is equal to 1 here, and that $w$ is scaled by $\mathrm{i}$ in our formulation), it is
\begin{align}\label{e:Pel}
w \to &-2 \mathrm{i} a_{+} + 2 \mathrm{i} a_{+}\tilde{z}\left(\lambda_1 \eps^{1/2} + \lambda_2 \eps + \ldots\right) \quad \mathrm{as} \quad \tilde{z} \to \infty, \, \eps \to 0,\\
w \to &-2 \mathrm{i} a_{-} - 2 \mathrm{i} a_{-}\tilde{z}\left(\lambda_1 \eps^{1/2} + \lambda_2 \eps + \ldots\right) \quad \mathrm{as} \quad \tilde{z} \to -\infty, \, \eps \to 0.\label{e:Pel2}
\end{align}
where $a_{\pm}$ are constants in $\tilde{z}$ that can scale with $\eps$. We found
\begin{equation}
    w \sim \frac{\mathrm{i}\eps^{1/2}\lambda_1}{\sqrt{2}}\left(1 + b_{11}\phi_0\right) + \mathrm{i}\eps\left(\frac{\lambda^{(0)}_2}{\sqrt{2}}  + b_{21}\phi_0 + b_{22}(\sqrt{2} - \tilde{z}\phi_0)\right) + \ldots \quad \mathrm{as} \quad \eps \to 0,
\end{equation}
where $\lambda_1 \sim \lambda_1^{(0)} + h \lambda_1^{(1)} + \ldots$. We require the asymptotic behaviour as $\tilde{z} \to \pm\infty$ to balance against \eqref{e:Pel}, which is 
\begin{align}\label{e:mylimit}
    w \sim \,\,&\mathrm{i}\eps^{1/2}\left(\frac{\lambda_1}{\sqrt{2}}+b_{11}\right) + \mathrm{i}\eps\left(\frac{\lambda_2}{\sqrt{2}}  + b_{21} + \sqrt{2}b_{22}- b_{22}\tilde{z}\right) + \ldots \quad \mathrm{as} \quad \tilde{z} \to \infty,\,\eps \to 0,\\
    w \sim \,\,&\mathrm{i}\eps^{1/2}\left(\frac{\lambda_1}{\sqrt{2}}-b_{11}\right)  + \mathrm{i}\eps\left(\frac{\lambda_2}{\sqrt{2}}  + b_{21} + \sqrt{2}b_{22}+ b_{22}\tilde{z}\right) + \ldots \quad \mathrm{as} \quad \tilde{z} \to \infty,\,\eps \to 0.\label{e:mylimit2}
\end{align}
Balancing the leading-order terms in \eqref{e:Pel}--\eqref{e:Pel2} and \eqref{e:mylimit}--\eqref{e:mylimit2} as $\eps \to 0$ gives
\begin{equation}\label{eq:apam}
    a_+ = -\frac{\sqrt{2}\lambda_1+2b_{11}}{4},\qquad a_- = -\frac{\sqrt{2}\lambda_1-2b_{11}}{4}.
\end{equation}
Now we balance the $\mathcal{O}(\tilde{z}\eps)$ terms of \eqref{e:Pel}--\eqref{e:Pel2} and \eqref{e:mylimit}--\eqref{e:mylimit2} in the dual limit to find 
\begin{equation}\label{e:b22}
 b_{11} = 0, \qquad b_{22} = \frac{\lambda_1^2}{2\sqrt{2}} \sim \frac{\left(\lambda_1^{(0)}\right)^2}{2\sqrt{2}} \quad \mathrm{as} \quad h \to 0. 
\end{equation}
Using \eqref{e:b22} in \eqref{e:a22old} gives
\begin{equation}
    a_{22} = \frac{\lambda_1^2}{8\sqrt{2}} \sim \frac{\left(\lambda^{(0)}_1\right)^2}{8\sqrt{2}}\quad \mathrm{as} \quad h \to 0. 
\end{equation} 
We have now determined all of the constants we require to complete the analysis. 

\subsubsection{Fredholm alternative}

Balancing terms of \eqref{eq:L-L+}, \eqref{lambda-exp}, and \eqref{v-w-exp}  at $\mathcal{O}(\eps^{3/2})$ gives
\begin{equation}
    L_+ v_3 = -\mathrm{i}\lambda_1 w_2 - \mathrm{i} \lambda_2 w_1.
\end{equation}
To ensure that the solution does not experience exponential growth in the limit that {$\tilde{z} \to \infty$}, this system must satisfy the Fredholm alternative. 
In this case, that implies
\begin{equation}
    -\mathrm{i}\lambda_1 \langle V_1, w_2 \rangle - \mathrm{i} \lambda_2 \langle V_1, w_1\rangle = 0.
\end{equation}
Evaluating this explicitly with $w_1$ in (\ref{w1explicitly}) and $w_2$ in (\ref{w2explicit}) gives
\begin{equation}
\label{lambda-2}
    \frac{\lambda_1\lambda_2}{2} + \frac{\lambda_1}{8}(\sqrt{2}\lambda_1^2 + 4\lambda_2) = 0,
\quad \Rightarrow \quad 
    \lambda_2 = -\frac{\lambda_1^2}{4\sqrt{2}}.
\end{equation}
Since $\lambda_1$ is imaginary, this quantity is real and the solution is unstable. Furthermore, we have been able to calculate the correction to the eigenvalue explicitly. For the onsite-centered solution, this gives the prediction for $\lambda_2$ 
\begin{equation}
    \lambda_2 = -\frac{\lambda_1^2}{4\sqrt{2}} \sim \frac{5\pi\Lambda}{\sqrt{2}h^5} \left(1 - \frac{\pi^2 h }{12\sqrt{2}}\right)^2 \sim \frac{5\pi^2\Lambda}{\sqrt{2}h^5} \left(1 - \frac{\pi^2 h }{6\sqrt{2}}\right) = \lambda_2^{(0)} + h \lambda_2^{(1)} \quad \mathrm{as} \quad h \to 0,
\end{equation}
From symmetry arguments, this in fact predicts four eigenvalues in total, described by
\begin{equation}
    \lambda \sim \pm\eps^{1/2}\lambda_1 \pm \eps\lambda_2 \quad \mathrm{as} \quad h \to 0,
\end{equation}
where the two sign choices are made independently, $\lambda_1^{(0)}$ and $\lambda_1^{(1)}$ are imaginary, $\lambda_2^{(0)}$ and $\lambda_2^{(1)}$ are real, and $\eps$ is a real scaling factor that is exponentially small in the limit $ h \to 0$. Hence, we find that the real part of the eigenvalues is
\begin{equation}
\label{lambda-2-onsite}
    \epsilon \lambda_2  \sim \pm\frac{5\pi^2\Lambda}{\sqrt{2}h^5} \left(1 - \frac{\pi^2 h }{6\sqrt{2}}\right)\mathrm{e}^{-\frac{\sqrt{2}\pi^2}{h}} \approx \pm(12508.376-14549.043h)h^{-5}\mathrm{e}^{-\frac{\sqrt{2}\pi^2}{h}} \quad \mathrm{as} \quad h \to 0.
\end{equation}
This means that even though the leading-order solution for $\lambda_1$ is imaginary for the onsite-centered solution, 
the real correction term for $\lambda_2$ ensures that the onsite-centered solution is unstable.

\subsection{Comparison of $\lambda_2$ with the pinning theory from \cite{PelinovskyKevrekidis2008}}

We show that the prediction of the correction $\lambda_2$ in \eqref{lambda-2} can be re-derived 
from the theory of the soliton pinning in an external potential in \cite{PelinovskyKevrekidis2008}. 
Moreover, this allows us to compute the effective Peierls--Nabarro potential due to the discrete lattice. 

We recall from \cite{PelinovskyKevrekidis2008} that the continuum NLS equation with an external potential
was considered in the form:
\begin{equation}
i u_t
=
-\frac{1}{2} u_{xx}
+
f(|u|^2)\,u
+
 \epsilon  V(x)\,u .
\label{1.2}
\end{equation}
It was shown in Theorems 2.12 and 4.11 of \cite{PelinovskyKevrekidis2008} that 
the dark soliton with the profile $\phi_0$ satisfying $|\phi_0(x)| \to 1$ as $|x| \to \infty$ 
are pinned as $\epsilon \to 0$ to simple zeros $s_0$ of the effective potential $M'(s)$ given by 
\begin{equation}
M'(s) = \int_{\mathbb{R}} V'(x)\,
\bigl[ 1-\phi_0^2(x-s)\bigr]\,
dx,
\label{1.5}
\end{equation}
while $M''(s_0)$ determines eigenvalues $\lambda$ of the spectral stability problem 
according to the characteristic equation 
\begin{equation}
\mathrm{Re}\,\lambda > 0 :
\qquad
\left(P'\big|_{v\downarrow 0}\right) \lambda^{2}
-
 \epsilon \,
\frac{\left(S'\big|_{v\downarrow 0}\right)^2 M''(s_0)}
{2 \sqrt{f'(1)} \left(P'\big|_{v\downarrow 0}\right)}
\,\lambda
+
 \epsilon \, M''(s_0)
=
\mathcal{O}( \epsilon ^{2}),
\label{4.15}
\end{equation}
where $P[u]$ and $S[u]$ were the conserved (renormalized) momentum and the conserved phase difference defined as
\begin{equation}
P[u]
=
\frac{i}{2}
\int_{\mathbb{R}}
\left(\bar u u_x - u \bar u_x\right)
\left(1-\frac{1}{|u|^2}\right)
dx 
\label{3.4}
\end{equation}
and
\begin{equation}
S[u]
=
\frac{i}{2}
\int_{\mathbb{R}}
\left(
\frac{\bar u_x}{\bar u}
-
\frac{u_x}{u}
\right)
dx.
\label{3.6}
\end{equation}
These conserved quantities were computed on the family of traveling dark soliton $u = \phi(x-vt)$ parametrized by 
the wave speed $v$ and the prime stands for the derivative with respect to $v$ as $v \downarrow 0$. 
In the most interesting case of $f(s)=s$,
the exact traveling dark solitons (up to the $e^{-i t}$ phase factor) exists in the form:
\begin{equation}
\phi(x - vt) = k \tanh(k (x-v t)) + i v,
\qquad
k = \sqrt{1-v^2},
\label{2.18}
\end{equation}

To implement these results to the case of our studies, we notice the $1/2$ factor mismatch in front
of the Laplacian term. Hence for our setting, we need to rescale $x$ and $v$ by $\frac{1}{\sqrt{2}}$ 
and replace the solutions of Eq.~(\ref{2.18}) to the form:
\begin{equation}
\phi(x - vt) = k \tanh(\frac{k}{\sqrt{2}} (x- v t)) + i \frac{v}{\sqrt{2}},
\qquad
k = \sqrt{1-\frac{v^2}{2}}.
\label{2.18b}
\end{equation}
On the other hand, the characteristic equation (\ref{4.15}) and the conserved quantities (\ref{3.4}) and (\ref{3.6}) remain 
in the same form. Using the exact solution~\eqref{2.18b}, we obtain 
\begin{equation}
P'\big|_{v\downarrow 0} = 2\sqrt{2},
\qquad
S'\big|_{v\downarrow 0} = \sqrt{2},
\end{equation}
so that the characteristic equation (\ref{4.15}) can be written in the form:
\begin{equation}
\mathrm{Re}\,\lambda > 0 :
\qquad
\lambda^{2}
+
\frac{ \epsilon }{2 \sqrt{2}} M''(s_0)
\left(
1 - \frac{\lambda}{2 \sqrt{2}}
\right)
=
\mathcal{O}( \epsilon ^{2}).
\label{4.24}
\end{equation}
Expanding like in (\ref{lambda-exp}),
$$
\lambda = \epsilon^{1/2} \lambda_1 + \epsilon \lambda_2 + \mathcal{O}(\epsilon^{3/2}), 
$$
we obtain from (\ref{4.24}) that 
\begin{equation}
    \label{4.24-impr}
\lambda_1^2 + \frac{1}{2 \sqrt{2}} M''(s_0) = 0, \quad \lambda_2 = \frac{1}{16} M''(s_0) = -\frac{\lambda_1^2}{4 \sqrt{2}},
\end{equation}
where the second relation coincides with (\ref{lambda-2}). By using (\ref{lambda-1}) and the first relation, we recover the 
derivative of the effective potential as 
\begin{equation}
    \label{M-two-derivatives}
M''(s_0) \sim \frac{40 \sqrt{2} \pi^2 \Lambda}{ h ^5}\mathrm{e}^{-\frac{\sqrt{2}\pi^2}{h} } \cos(2\pi s_0).
\end{equation}
Integration recovers the effective Peierls--Nabarro potential 
$$
M'(s) \sim \frac{20 \sqrt{2} \pi \Lambda}{ h ^5}\mathrm{e}^{-\frac{sqrt{2}\pi^2}{h} } \sin(2\pi s),
$$
which replaces the effective continuous potential (\ref{1.5}) in the discrete setting.

\section{Numerical Results}

Before presenting our numerical results,
we briefly summarize the currently available predictions regarding the eigenvalues associated
with the intersite and onsite solitary waves. 

We start with the simpler intersite case which
was considered theoretically near the anti-continuum limit
in~\cite{Pelinovsky_2008} and near the continuum one of in~\cite{ADRIANO2025134848}.
In this case, the discrete solitary wave is
well-known (already from the seminal work
of~\cite{johkiv}) to possess a real eigenvalue pair in the corresponding
linearization.
This pair is known to bifurcate from the
origin of the spectral plane along the real axis {\it already in the vicinity of the
anti-continuum limit} (i.e., for lattice spacing
$h \rightarrow \infty$)  \cite{Pelinovsky_2008}. The associated pair
{\it remains real for all $h$} and finally approaches
the origin to ``restore'' translational invariance
through an exponentially small dependence 
as $h \rightarrow 0$ \cite{ADRIANO2025134848}.
Revisiting relevant quantitative details,
we recall that the work of~\cite{ADRIANO2025134848}
found the relevant eigenvalue to be
{\begin{eqnarray}
    \lambda \approx \pm \sqrt{4 \sqrt{2} \pi^2 2534.73} 
    h^{-5/2} e^{-\frac{\sqrt{2} \pi^2}{2 h}},
    \label{pred1}
    \end{eqnarray}
where, to enable direct comparison, we have continued the calculation of the constant from~\cite{ADRIANO2025134848} 
to have the equivalent precision to our calculation of the unstable eigenvalue $\lambda$.} The numerical computations of~\cite{KusdiantaraAdrianoSusanto2025} deemed
this prediction to bear the correct functional
dependence on $h$, yet miss the leading order
prefactor. The best fit numerical approximation,
motivated by the corresponding expression
for bright discrete solitary waves in~\cite{LustriKevrekidisChapman2025},
was found to be:
\begin{eqnarray}
    \lambda \approx \pm \left(265.238 - 135.983 h
    \right)   h^{-5/2} e^{-\frac{\sqrt{2} \pi^2}{2 h}}
    \label{pred2}
\end{eqnarray}
Our predictions of the unstable eigenvalue $\lambda$ for the intersite solution, based on~\cite{PelinovskyKevrekidis2008},
are found from (\ref{4.24}) and (\ref{M-two-derivatives}), 
\begin{eqnarray}
    \lambda \approx \pm \sqrt{2\sqrt{2} \pi^2 2534.73 
    h^{-5} e^{-\frac{\sqrt{2} \pi^2}{h}}}
    = 266.004 h^{-5/2} e^{-\frac{\sqrt{2} \pi^2}{2 h}}
\label{pred5}
\end{eqnarray}
which is in {\it excellent agreement} with the
leading order of the multiprecision numerical findings
of~\cite{KusdiantaraAdrianoSusanto2025}. 
The reason for the discrepancy in (\ref{pred1}) from \cite{ADRIANO2025134848}, compared 
to the correct prediction (\ref{pred5}), is due to the choice in solving the linear 
inhomogeneous equation (\ref{w-1-eq}) for the correction term $w_1$. The solution 
was chosen to be unbounded in \cite{ADRIANO2025134848}, whereas the relevant solution must be bounded 
as in our Eq. (\ref{w1explicitly}) due to $b_{12} = 0$. The correct prediction (\ref{pred5}) 
is obtained from the bounded solution for $w_1$ independently on the value of $b_{11}$.

In the case of onsite solution, the binomial of (\ref{4.24}), (\ref{4.24-impr}), and (\ref{M-two-derivatives})
yields (again up to exponentially small corrections):
\begin{eqnarray}
    \lambda \approx 
    \pm i 266.004 h^{-5/2} e^{-\frac{\sqrt{2} \pi^2}{2 h}} \pm 12508.376 h^{-5} e^{-\frac{\sqrt{2} \pi^2}{h}}
    \label{pred6}
\end{eqnarray}
The work of~\cite{KusdiantaraAdrianoSusanto2025}
provides a fit only in a neighborhood far from 
$h \rightarrow 0$, hence we do not compare (\ref{pred6}) to \cite{KusdiantaraAdrianoSusanto2025}.

Our {\it full} exponential asymptotics prediction
for the case of intersite soluton, incorporating the
next order correction to the leading order, is given by (\ref{lambda-intersite}) and thus improving
upon Eq.~(\ref{pred5}) reads:
\begin{eqnarray}
 \Lambda \approx 
    \pm    (266.004 -154.7 h)
    h^{-5/2} e^{-\frac{\sqrt{2} \pi^2}{2 h}} 
    \label{pred7}
\end{eqnarray}
which can be seen to be highly proximal {---in its leading order prediction---}
to the multiprecision numerical fit of~\cite{KusdiantaraAdrianoSusanto2025}.
{It is also worthwhile to note how our next order correction
improves upon the relevant prediction in comparison to the numerical fit, 
as the latter was performed with data at some distance
from the $h\rightarrow 0$ limit}.

The summary of all the analytical and numerical results 
of earlier studies~\cite{ADRIANO2025134848,KusdiantaraAdrianoSusanto2025,PelinovskyKevrekidis2008} culminating into this work are summarized in Table~\ref{tab:IS_eigenvalue_comparison}. The table illustrates
at a glance how the present work's analysis amends the prediction
of~\cite{ADRIANO2025134848}, extends the leading order
adaptation (to the present discrete setting) of~\cite{PelinovskyKevrekidis2008}, while matching very
closely the best fit numerical findings of~\cite{KusdiantaraAdrianoSusanto2025}.

\begin{table}[h!]
\centering
\caption{Comparison of Different Predictions for the IS Discrete Soliton Positive Real Eigenvalue}
\label{tab:IS_eigenvalue_comparison}
\begin{tabular}{|c|c|c|}
\hline
Numerical Prefactor & Reference Source & Origin \\
\hline
${\sqrt{4 \sqrt{2} \pi^2 2534.73}}$ & \cite{ADRIANO2025134848} & Analytical \\
\hline
$265.238 - 135.983\,h$ & \cite{KusdiantaraAdrianoSusanto2025} & Numerical Fit \\
\hline
${\sqrt{2\sqrt{2}\,\pi^2\,2534.73} \equiv 266.004}$ & \cite{PelinovskyKevrekidis2008} & Analytical \\
\hline
$266.004 - 154.7\,h$ & Present Work & Analytical \\
\hline
\end{tabular}
\end{table}

\begin{figure}[htbp]
  \centering
  \begin{subfigure}{0.49\textwidth}
    \centering
    \includegraphics[width=\linewidth]{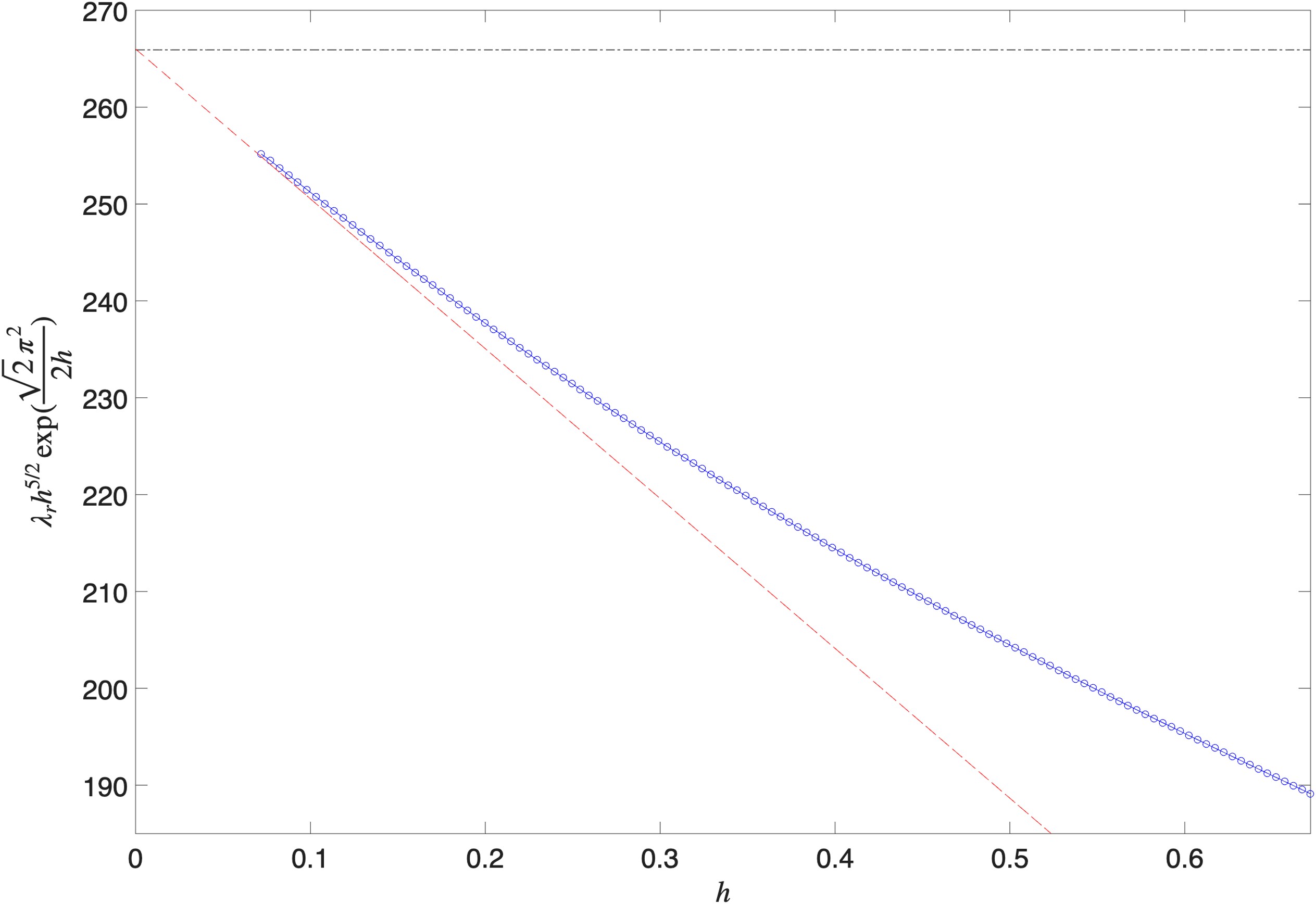}
  \end{subfigure}\hfill
  \begin{subfigure}{0.49\textwidth}
    \centering
    \includegraphics[width=\linewidth]{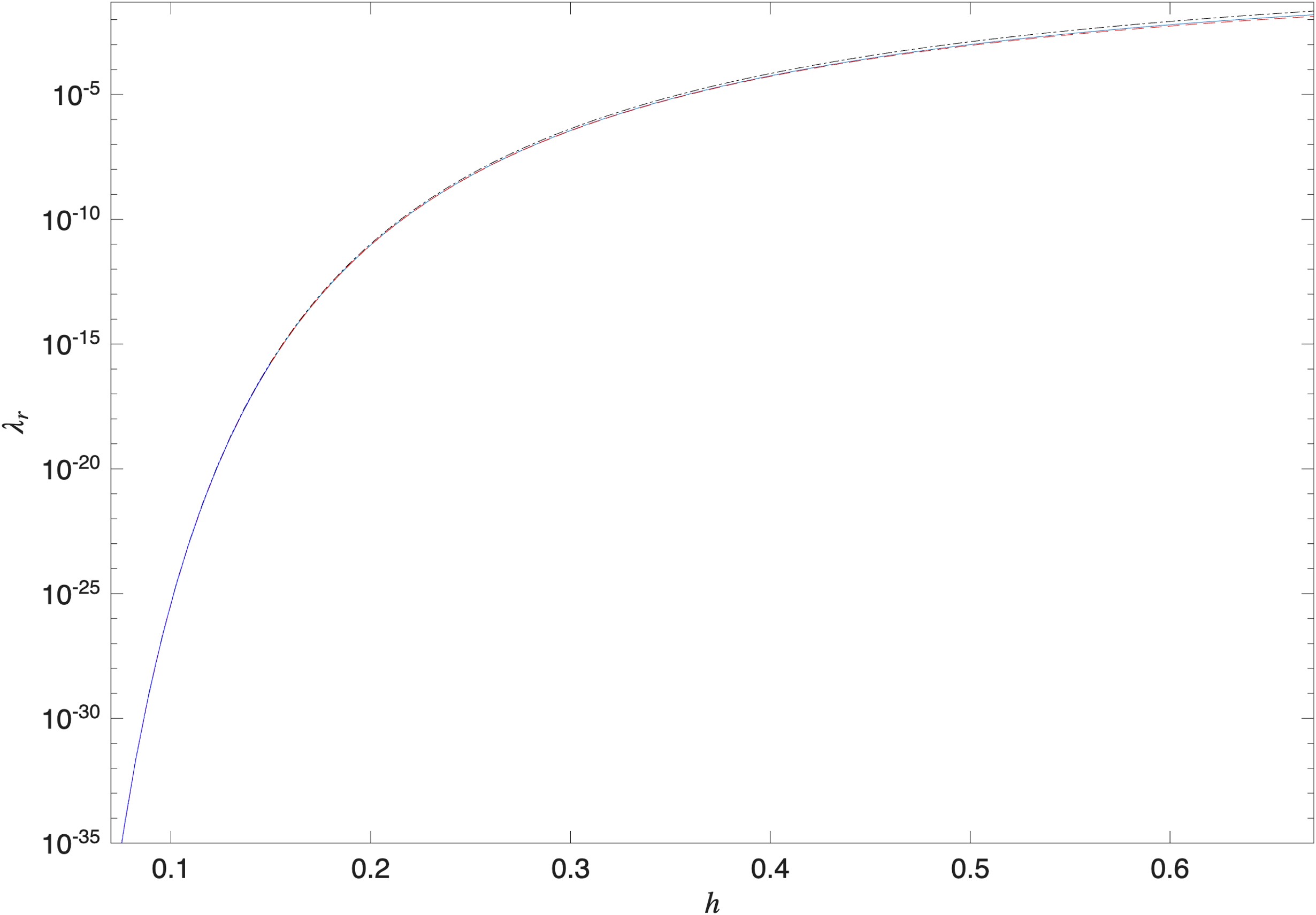}
  \end{subfigure}
  \caption{The eigenvalue of the intersite solution is shown both upon its multiplication
  by $h^{5/2} e^{\frac{\sqrt{2} \pi^2}{2 h}}$
  (which renders its asymptotic form transparent)
  in the left panel and ``as is'' in the right 
  panel. The numerical data are shown in blue,
  while the red dashed provides the
  exponential asymptotic prediction 
  of Eq.~\eqref{pred7} and the dash-dotted black
  curve yields the leading order approximation
  of~Eq.~(\ref{pred5}).}
  \label{fig:numerical-is2-is3}
\end{figure}

The numerical prediction of Eq.~(\ref{pred7}) is
given by the red dashed line in both panels of
Fig.~\ref{fig:numerical-is2-is3}. The 
leading order result (\ref{pred5}) is captured
by the black dash-dotted line. The latter
can be seen (through a suitable magnification)
to start deviating from the numerical results
for $h \approx 0.2$. Remarkably, the
prediction of Eq.~(\ref{pred7}) cannot be really
distinguished from the numerics over the
scale of the graph, i.e., even up to $h \approx 0.7$
and quite remarkably for about 35 orders
of magnitude of the variation of the relevant
real eigenvalue (see the right panel).
The left panel multiplies the eigenvalue by both
the power and the exponential, lucidly illustrating
the approach to the asymptotic prediction 
as $h \rightarrow 0$. These results were obtained
with computations using the 
Multiprecision Computing Toolbox within Matlab
(but with smaller lattices involving $1000$ nodes; cf.
with the larger lattice computations below).

\begin{comment}
\begin{equation}
\lambda^2
=
\frac{ h }{\mu_0}
\left(
M''(s_0) +  h  \lambda_0 \lambda
\right)
+ \mathcal{O}( h ^2).
\label{4.15}
\end{equation}

\begin{equation}
\lambda^2
+
 h  \lambda_0 \lambda
-
\frac{ h }{\mu_0} M''(s_0)
=
0 .
\tag{4.24}
\end{equation}
\end{comment}

% ------------------------------------------------------------------------

\begin{figure}[htbp]
  \centering
  \begin{subfigure}{0.49\textwidth}
    \centering
    \includegraphics[width=\linewidth]{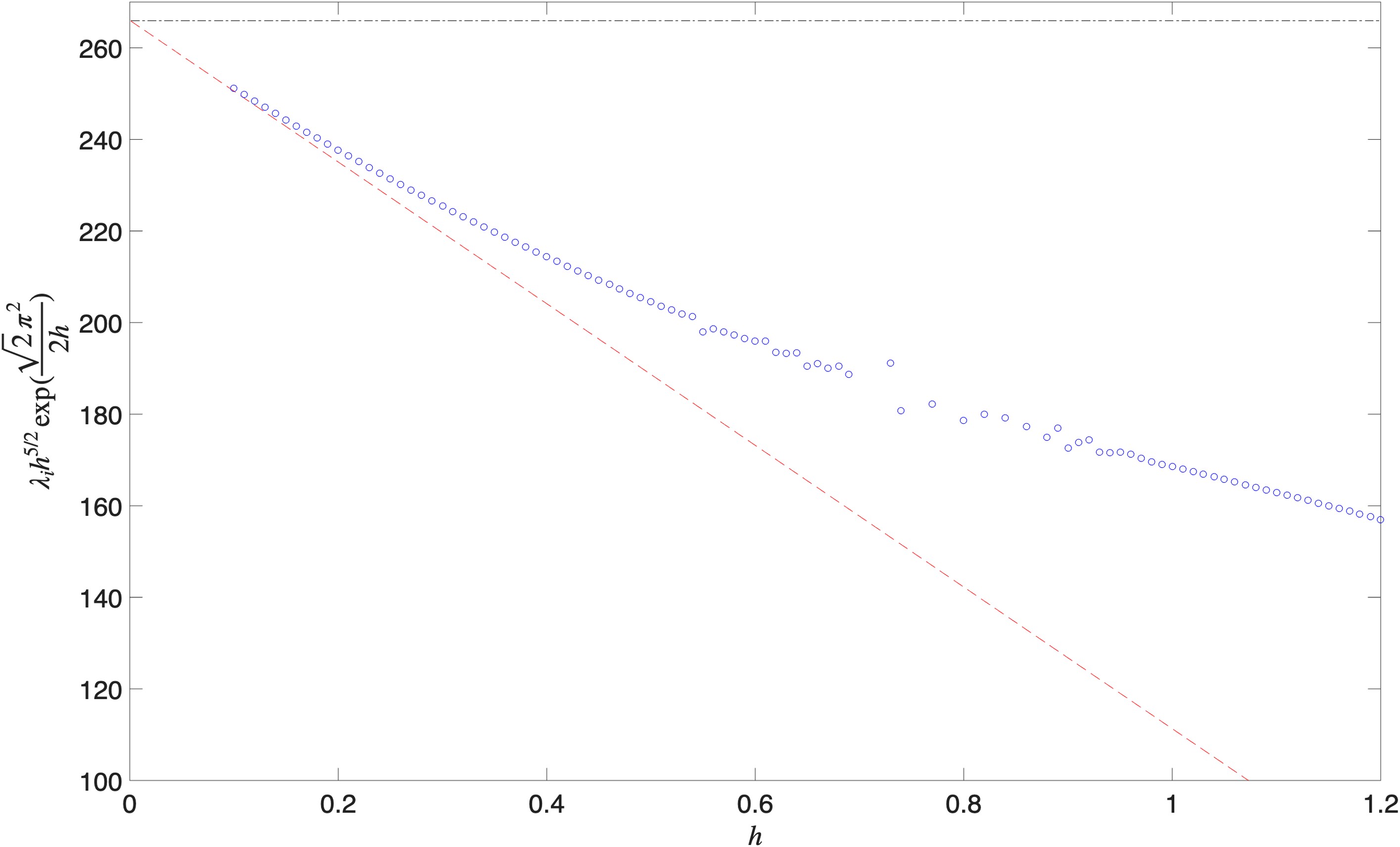}
  \end{subfigure}\hfill
  \begin{subfigure}{0.49\textwidth}
    \centering
    \includegraphics[width=\linewidth]{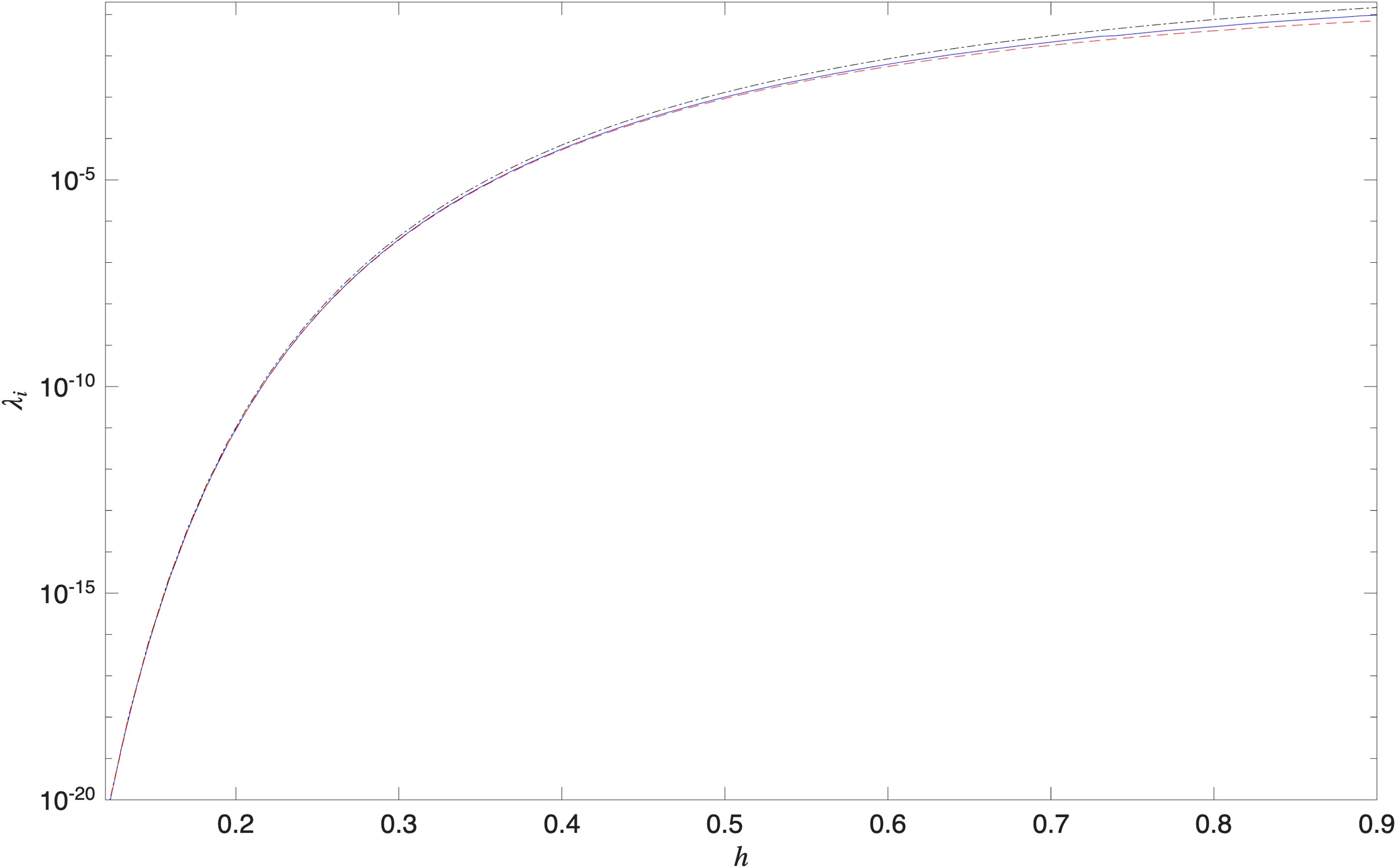}
  \end{subfigure}
  \caption{Similar results to Figure \ref{fig:numerical-is2-is3}
  but for the imaginary part of the complex
  eigenvalue associated with the onsite
 solution. The prediction is essentially 
  the same as above (now on the imaginary axis)
  and the agreement of both the leading
  order asymptotics (\ref{pred6}) and of the next-order correction (\ref{pred6-impr}) 
  documented again by the black, dash-dotted
  and the red dashed curves, respectively.}
  \label{fig:numerical-2d-2c}
\end{figure}

Figure~\ref{fig:numerical-2d-2c} presents similar results to the ones
above but for the onsite solution. Here, a clarification needs to be made
(which is partially aligned with the explanations of~\cite{KusdiantaraAdrianoSusanto2025}).
In order to use the Multiprecision Computing Toolbox, we found it necessary to
restrict our considerations to computations with $1000$ nodes. In that case, 
the continuous spectrum becomes discretized. Accordingly, as $h \rightarrow 0$,
the complex quartet associated with the onsite mode instability traces gaps
between the discretized eigenvalues and accordingly returns to the imaginary
axis. Similar findings have been reported in~\cite{johkiv} (see Fig.~2 therein
and the corresponding ``arcs'' of eigenvalues jumping in and out of the imaginary
axis), as well as in~\cite{PelinovskyKevrekidis2008} (see, e.g., Fig.~4 of that work,
and the associated non-monotonic dips therein). As the continuum limit is approached
the final ``crossing'' of this sort takes place around $h \approx 0.8$, hence the
``scatter'' of eigenvalues near that neighborhood, as observed in the left
panel of Fig.~\ref{fig:numerical-2d-2c}. For lower values of $h$ and as the 
continuum limit is approached, the numerical problem, due to its finite size,
has the translational eigenvalue approaching the origin as purely imaginary.
Nevertheless, and remarkably so, this imaginary approach is found to be 
excellently aligned with the theoretical prediction for the imaginary part of the
eigenvalue as given in Eq.~(\ref{pred6}) to leading order. In fact, 
with the next-order correction to the leading order given by (\ref{lambda-onsite}) and (\ref{lambda-2-onsite}), 
we have 
\begin{eqnarray}
    \lambda \approx 
    \pm i (266.004 - 154.700 h) h^{-5/2} e^{-\frac{\sqrt{2} \pi^2}{2 h}} \pm 
    (12508.376 - 14549.043 h) h^{-5} e^{-\frac{\sqrt{2} \pi^2}{h}}.
    \label{pred6-impr}
\end{eqnarray}
The Multiprecision Computing Toolbox enables capturing this prediction 
very accurately to several orders of magnitude as can be seen in the right panel
of Figure~\ref{fig:numerical-2d-2c}. On the other hand, the leading order prediction can again be visibly
discerned to start gradually deviating around $h \approx 0.2$. Nevertheless,
the dominant prediction can be deemed to be extremely accurate up to $h \approx 0.2$
and the next order correction, similarly, even beyond $h \approx 0.5$.

\begin{figure}[htbp]
  \centering
  \includegraphics[width=0.8\linewidth]{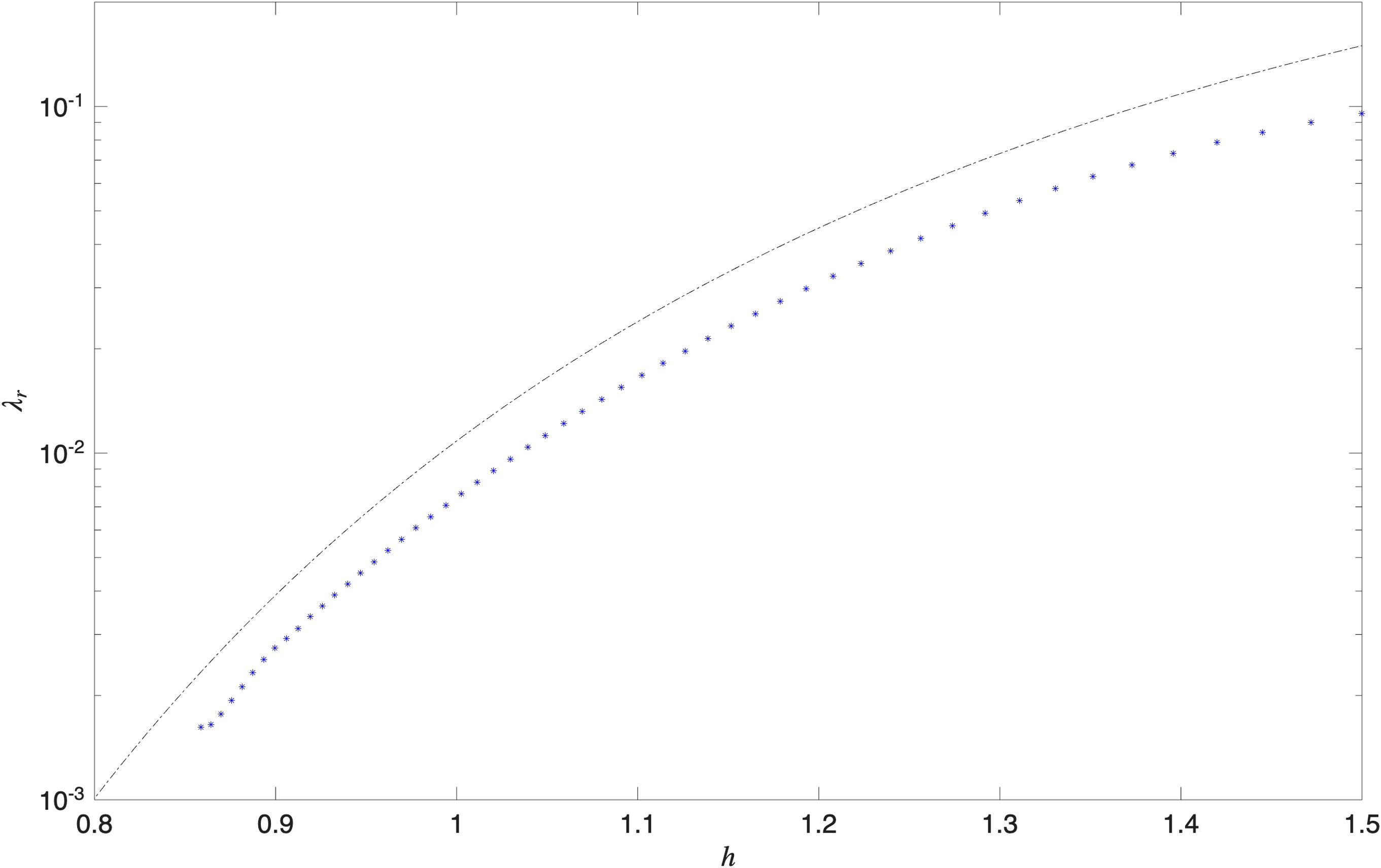}
  \caption{Numerical results for the real
  (unstable) part of the complex eigenvalue of
  the onsite configuration. The leading-order
  asymptotics of Eq.~(\ref{pred6}) is given
  by the black dash-dotted curve.}
  \label{fig:numerical-2f}
\end{figure}

The above considerations clearly underscore the difficulty of
capturing the real part of the eigenvalue quartet all the way to
$h \rightarrow 0$, a difficulty that forced the authors of~\cite{KusdiantaraAdrianoSusanto2025}
to perform a fit only up to $h \approx 0.9$. The real part of the
corresponding quartet as computed in much larger lattices of $3000$ nodes
(but without the Multiprecision Computing Toolbox, as the latter
computation proved to be infeasible with our computational 
resources)
is shown Fig.~\ref{fig:numerical-2f} down to $h \approx 0.9$. The 
leading-order prediction of Eq.~(\ref{pred6}) is compared to the corresponding
numerical result in the figure, with the deviation being comparable to what
one can observe in the right panel of Fig.~\ref{fig:numerical-2d-2c} for
the same parameter range. While we have every expectation that 
the leading-order prediction provided matches very well the numerical,
infinite lattice result as $h \rightarrow 0$, our computational
capabilities do not allow a full confirmation of that fact at present.
%Nevertheless, we believe that we have corroborated the feature that the predictions herein represent the most definitive (to date) approximation of the relevant translational eigenvalues via refined variants of exponential asymptotics.

\section{Conclusions and Future Challenges}

In the present work we have revisited a problem with considerable
history, as well as with very broad applicability. This concerns
the implications of discreteness and its associated breaking
of translational invariance in connection with the spectral features
of the multiple (intersite and onsite) dark discrete solitary waves that emerge through
the discretization of the model as potential stationary states.
In line with earlier studies, the breaking of translational invariance
leads to beyond-all-algebraic order (indeed exponentially small) eigenvalues,
modulated by a power of $h^{-\alpha}$, where $h$ is the lattice spacing.
While early studies such as~\cite{johaub} and~\cite{Todd_Kapitula_2001} captured the numerical
features of the bright solitary wave spectral analysis, it was
not until recently that the focusing problem was fully analyzed
in terms of the leading order~\cite{ADRIANO2025134848,LustriKevrekidisChapman2025} and
even its next order correction~\cite{LustriKevrekidisChapman2025}
for the eigenvalue associated with translation.
Nevertheless, the technical complexity of the dark solitary
waves of the defocusing DNLS, already outlined in the seminal
work of~\cite{johkiv}, have meant that the latter case has remained
elusive for a quarter of a century with recent efforts~\cite{ADRIANO2025134848,KusdiantaraAdrianoSusanto2025}
improving on the relevant numerics, yet not providing a conclusive
theoretical answer. The present work has been able to achieve that
{\it both} for the intersite dark solitary waves (and their
real eigenvalue pair) and for the considerably more elaborate
onsite dark solitary waves (bearing a complex eigenvalue quartet
near the continuum limit). In the process, we have also established
a firm connection with the analysis of pinning of the dark solitary waves in 
~\cite{PelinovskyKevrekidis2008} in the presence of external potentials. This connection may
bear further fruit towards describing the motion of the 
discrete solitary waves, by analogy to the corresponding 
equations of motion obtained in~\cite{PelinovskyKevrekidis2008}.

In addition to this dynamical vein for defocusing DNLS dark solitons,
we believe that this line of asymptotic efforts may lead to numerous further
developments. For instance, a topic that has received considerable
attention not only in the context of the DNLS~\cite{johaub}, but also
in other settings such as the discrete sine-Gordon~\cite{Peypel,Todd_Kapitula_2001} has been the study of 
internal modes that bifurcate from the continuous spectrum, leading
associated resonances ---sitting at the edge of the spectral band---
to transform into point spectrum eigenvalues. Although such modes were 
originally deemed to be power law in their bifurcations~\cite{Peypel,KevJon}, it was
subsequently discovered that they bear both exponential and power law
characteristics~\cite{KapKevJon,Todd_Kapitula_2001}. Nevertheless,
associated eigenvalue dependences on the lattice spacing never
reached the level of accuracy of recent predictions based on exponential
asymptotics, leaving an important open problem. 

An additional equally intriguing vein concerns exploring models
of either generalized power or of generalized dimension in the context
of the DNLS~\cite{MALOMED199691,JOHANSSON1998115}, as these models
possess stability variations (even for onsite branches) and invervals
of bistability. It would be interesting to explore how these models
approach the continuum limit as a function of the dimension $d$ or
the exponent $p$ of the nonlinearity and to explore how exponential
asymptotics may be able to capture the dependence not only on the
lattice spacing but also on such as an additional parameter such as $p$.

Finally, recent work has explored the setting of the so-called ``Maxwell
fronts'', i.e., kinks that arise from the competition of two different
nonlinearities, such as, e.g., quadratic and cubic, or cubic and quintic.
These kinks have been demonstrated to possess intriguing 
stability properties not only in one dimension~\cite{holmer2025orbitalstabilitykinksnls}, 
but also in higher dimensions where they are transversely stable~\cite{kinks_transverse}.
The work of~\cite{hadi_recent} has explored such kinks in the realm of
the lattice where they were shown to be continued from the anti-continuum all the way
to the continuum limit. A further characterization of their stability
properties based on the methods proposed herein is a particularly interesting subject
for future work.
%Such studies are currently in progress and will be reported in future
%publications.

\vspace{0.2cm}

{\bf Acknowledgements.} CJL acknowledges support from Australian Research Council Discovery Project DP240101666. 
This research was supported by the U.S. National Science Foundation under the award PHY-2408988 (PGK). 
This research was partly conducted while P.G.K. was  visiting the Okinawa Institute of Science and
Technology (OIST) through the Theoretical Sciences Visiting Program (TSVP), the University of
Sydney through the visitor program of the Sydney Mathematical Research Institute (SMRI) and the Department of Mechanical Engineering at Seoul National 
University through a Fulbright Fellowship. Their support is gratefully acknowledged.
Finally, this work was also  supported by a grant from the Simons Foundation [SFI-MPS-SFM-00011048, P.G.K]. 

\bibliographystyle{plain}
\bibliography{reference2}

\begin{thebibliography}{10}

\bibitem{AblowitzPrinariTrubatch}
M.~J. Ablowitz, B.~Prinari, and A.~D. Trubatch.
\newblock {\em Discrete and Continuous Nonlinear {S}chr\"odinger Systems}.
\newblock Cambridge University Press, Cambridge, 2004.

\bibitem{ADRIANO2025134848}
F.~T. Adriano, A.~N. Hasmi, R.~Kusdiantara, and H.~Susanto.
\newblock Exponential asymptotics of dark and bright solitons in the discrete
  nonlinear schrödinger equation.
\newblock {\em Physica D}, 481:134848, 2025.

\bibitem{hadi_recent}
F.~T. Adriano and H.~Susanto.
\newblock Maxwell fronts in the discrete nonlinear {S}chr{\"o}dinger equations
  with competing nonlinearities.
\newblock {\em Stud. Appl. Math.}, 156(2):e70191, 2026.

\bibitem{aniceto2019primer}
I.~Aniceto, G.~Ba{\c{s}}ar, and R.~Schiappa.
\newblock A primer on resurgent transseries and their asymptotics.
\newblock {\em Phys. Rep.}, 809:1--135, 2019.

\bibitem{chapman2005exponential}
S.~J. Chapman and D.~B. Mortimer.
\newblock Exponential asymptotics and {S}tokes lines in a partial differential
  equation.
\newblock {\em Proc. Roy. Soc. Lond. {A}}, 461(2060):2385--2421, 2005.

\bibitem{cretegny}
T.~Cretegny and S.~Aubry.
\newblock Spatially inhomogeneous time-periodic propagating waves in anharmonic
  systems.
\newblock {\em Phys. Rev. B}, 55:R11929--R11932, May 1997.

\bibitem{chriseil}
J.~C. Eilbeck and M.~Johansson.
\newblock The discrete nonlinear {S}chr\"odinger equation - 20 years on.
\newblock {\em Localization and Energy Transfer in Nonlinear Systems}, pages
  pp. 44--67, 2003.

\bibitem{Flach2008}
S.~Flach and A.~V. Gorbach.
\newblock {Discrete breathers — Advances in theory and applications}.
\newblock {\em Phys. Rep.}, 467(1-3):1--116, 2008.

\bibitem{Flach1999}
S.~Flach and K.~Kladko.
\newblock Moving discrete breathers?
\newblock {\em Physica D}, 127(1):61 -- 72, 1999.

\bibitem{GOMEZGARDENES2004213}
J.~Gómez-Gardeñes, F.~Falo, and L.~M. Floría.
\newblock Mobile localization in nonlinear {S}chr\"odinger lattices.
\newblock {\em Phys. Lett. A}, 332(3):213--219, 2004.

\bibitem{holmer2025orbitalstabilitykinksnls}
J.~Holmer, P.~G. Kevrekidis, and D.~E. Pelinovsky.
\newblock Orbital stability of kinks in the nls equation with competing
  nonlinearities, 2025.

\bibitem{johaub}
M.~Johansson and S.~Aubry.
\newblock Growth and decay of discrete nonlinear {S}chr\"odinger breathers
  interacting with internal modes or standing-wave phonons.
\newblock {\em Phys. Rev. E}, 61:5864--5879, May 2000.

\bibitem{JOHANSSON1998115}
M.~Johansson, S.~Aubry, Y.~B. Gaididei, P.~L. Christiansen, and K.~{\O}.
  Rasmussen.
\newblock Dynamics of breathers in discrete nonlinear {S}chr\"odinger models.
\newblock {\em Physica D}, 119(1):115--124, 1998.

\bibitem{johkiv}
M.~Johansson and Y.~S. Kivshar.
\newblock Discreteness-induced oscillatory instabilities of dark solitons.
\newblock {\em Phys. Rev. Lett.}, 82:85--88, Jan 1999.

\bibitem{Todd_Kapitula_2001}
T.~Kapitula and P.~G. Kevrekidis.
\newblock Stability of waves in discrete systems.
\newblock {\em Nonlinearity}, 14(3):533, may 2001.

\bibitem{KapKevJon}
T.~Kapitula, P.~G. Kevrekidis, and C.~K. R.~T. Jones.
\newblock Soliton internal mode bifurcations: Pure power law?
\newblock {\em Phys. Rev. E}, 63:036602, Feb 2001.

\bibitem{kev09}
P.~G. Kevrekidis.
\newblock {\em {The {D}iscrete {N}onlinear {S}chr{\"o}dinger {E}quation}}.
\newblock Springer-Verlag, Heidelberg, 1st edition, 2009.

\bibitem{KEVREKIDIS2012982}
P.~G. Kevrekidis, G.~J. Herring, S.~Lafortune, and Q.~E. Hoq.
\newblock The higher-dimensional {A}blowitz–{L}adik model: From
  (non-)integrability and solitary waves to surprising collapse properties and
  more exotic solutions.
\newblock {\em Phys. Lett. A}, 376(8):982--986, 2012.

\bibitem{KevJon}
P.~G. Kevrekidis and C.~K. R.~T. Jones.
\newblock Bifurcation of internal solitary wave modes from the essential
  spectrum.
\newblock {\em Phys. Rev. E}, 61:3114--3121, Mar 2000.

\bibitem{king_chapman_2001}
J.~R. King and S.~J. Chapman.
\newblock Asymptotics beyond all orders and {S}tokes lines in nonlinear
  differential-difference equations.
\newblock {\em Eur. J. Appl. Math.}, 12(4):433–463, 2001.

\bibitem{Peypel}
Y.~S. Kivshar, D.~E. Pelinovsky, T.~Cretegny, and M.~Peyrard.
\newblock Internal modes of solitary waves.
\newblock {\em Phys. Rev. Lett.}, 80:5032--5035, Jun 1998.

\bibitem{campbell}
Yuri~S. Kivshar and David~K. Campbell.
\newblock Peierls-{N}abarro potential barrier for highly localized nonlinear
  modes.
\newblock {\em Phys. Rev. E}, 48:3077--3081, Oct 1993.

\bibitem{KusdiantaraAdrianoSusanto2025}
R.~Kusdiantara, F.~T. Adriano, and H.~Susanto.
\newblock Multiprecision computation of bright and dark solitons in the
  discrete nonlinear {S}chr\"odinger equation.
\newblock Preprint, 2025.

\bibitem{lustri2025borelpadeexponentialasymptoticsdiscrete}
C.~J. Lustri, I.~Aniceto, and P.~G. Kevrekidis.
\newblock Borel-{P}ad\'e exponential asymptotics for the discrete nonlinear
  {S}chr{\"o}dinger model with next-to-nearest neighbour interactions.
\newblock {\em SIAM J. Appl. Math.}, 2026 (In press).

\bibitem{LustriKevrekidisChapman2025}
C.~J. Lustri, P.~G. Kevrekidis, and S.~Jonathan Chapman.
\newblock Exponential asymptotics for translational modes in the discrete
  nonlinear {S}chr\"odinger model.
\newblock {\em Quart. Appl. Math.}, 2025.
\newblock Published electronically June 5, 2025.

\bibitem{RSMacKay_1994}
R~S MacKay and S~Aubry.
\newblock Proof of existence of breathers for time-reversible or {H}amiltonian
  networks of weakly coupled oscillators.
\newblock {\em Nonlinearity}, 7(6):1623, nov 1994.

\bibitem{MALOMED199691}
B.~Malomed and M.~I. Weinstein.
\newblock Soliton dynamics in the discrete nonlinear {S}chr\"odinger equation.
\newblock {\em Phys. Lett. A}, 220(1):91--96, 1996.

\bibitem{DV}
B.~A. Malomed and P.~G. Kevrekidis.
\newblock Discrete vortex solitons.
\newblock {\em Phys. Rev. E}, 64:026601, Jul 2001.

\bibitem{kinks_transverse}
S.~I. Mistakidis, G.~Bougas, G.~C. Katsimiga, and P.~G. Kevrekidis.
\newblock Generic transverse stability of kink structures in atomic and optical
  nonlinear media with competing attractive and repulsive interactions.
\newblock {\em Phys. Rev. Lett.}, 134:123402, Mar 2025.

\bibitem{Daalhuis}
A.~B. {Olde Daalhuis}, S.~J. Chapman, J.~R. King, J.~R. Ockendon, and R.~H.
  Tew.
\newblock Stokes phenomenon and matched asymptotic expansions.
\newblock {\em SIAM J. Appl. Math.}, 55(6):1469--1483, 1995.

\bibitem{oxtoby}
O.~F. Oxtoby and I.~V. Barashenkov.
\newblock Moving solitons in the discrete nonlinear {S}chr\"odinger equation.
\newblock {\em Phys. Rev. E}, 76:036603, Sep 2007.

\bibitem{Pelinovsky_2006}
D.~E. Pelinovsky.
\newblock Translationally invariant nonlinear {S}chr\"odinger lattices.
\newblock {\em Nonlinearity}, 19(11):2695, oct 2006.

\bibitem{Pelinovsky_2011}
D.~E. Pelinovsky.
\newblock {\em Localization in Periodic Potentials: From Schrödinger Operators
  to the Gross–Pitaevskii Equation}.
\newblock London Mathematical Society Lecture Note Series. Cambridge University
  Press, 2011.

\bibitem{PelinovskyKevrekidis2008}
D.~E. Pelinovsky and P.~G. Kevrekidis.
\newblock Dark solitons in external potentials.
\newblock {\em Z. Angew. Math. Phys.}, 59:559--599, 2008.

\bibitem{Pelinovsky_2008}
D.~E. Pelinovsky and P.~G. Kevrekidis.
\newblock Stability of discrete dark solitons in nonlinear {S}chr{\"o}dinger
  lattices.
\newblock {\em J. Phys. A}, 41(18):185206, apr 2008.

\bibitem{PEYRARD198488}
M.~Peyrard and M.~D. Kruskal.
\newblock Kink dynamics in the highly discrete sine-{G}ordon system.
\newblock {\em Physica D}, 14(1):88--102, 1984.

\end{thebibliography}

\end{document}